\documentclass[preprint2]{aastex}		

\received{20 December 1999}
\revised{}
\accepted{}

\cpright{AAS}{2000}

\slugcomment{Submitted to ApJL, December 20, 1999}

\shortauthors{Hjorth et al.}
\shorttitle{GRB~990712 late decay and host galaxy}

\newcommand{\etal}{et~al.}

\newcommand{\hst}{{\sl HST\/}}

\newcommand{\paper}{{\sl Letter\/}}
\newcommand{\kms}{{\rm km\,s}^{-1}}
\newcommand{\ho}{\kms\ {\rm Mpc}^{-1}}

\begin{document}

\title{The late afterglow and host galaxy of GRB~990712
       {}\footnote{\rm Based on observations collected at the European
		 Southern Observatory, La Silla, Chile (ESO Programme
                 64.O--0019) and on observations with the NASA/ESA
                 {\sl Hubble Space Telescope}, obtained from the data
                 archive at the Space Telescope Science Institute, which
                 is operated by the Association of Universities for
                 Research in Astronomy, Inc.\ under NASA contract
                 NAS5-26555.}}

\author{Jens~Hjorth\altaffilmark{2},
Stephen~Holland\altaffilmark{3},
Frederic~Courbin\altaffilmark{4},
Arnon~Dar\altaffilmark{5},
Lisbeth~F.~Olsen\altaffilmark{2},
Marco~Scodeggio\altaffilmark{6,7}
}


\altaffiltext{2}{Astronomical Observatory, University of Copenhagen,
                 Juliane Maries Vej 30, DK--2100 Copenhagen \O, Denmark; 
                 {jens@astro.ku.dk}.}
\altaffiltext{3}{Institute of Physics and Astronomy, University of Aarhus,
                 DK--8000 {\AA}rhus C, Denmark}
\altaffiltext{4}{Department of Astronomy, P. Universidad Cat\'olica,
                 Casilla 306, Santiago, Chile}
\altaffiltext{5}{Department of Physics, Technion, Haifa 32000, Israel}
\altaffiltext{6}{European Southern Observatory, Karl-Schwarzschild-Str.~2,
		 D--85748, Garching bei M{\"u}nchen, Germany}
\altaffiltext{7}{Istituto di Fisica Cosmica ``G. Occhialini'',
                 via Bassini 15, I--20133 Milano, Italy}

\begin{abstract}

	We present deep \hst\ imaging, as well as ground-based imaging
and spectroscopy, of the optical afterglow associated with the
long-duration gamma-ray burst GRB~990712 and its host
galaxy. The data were obtained 48--123 days after the burst occurred.
The magnitudes of the host ($R=21.9$, $V=22.5$) and optical afterglow
($R=25.4$, $V=25.8$, 47.7 days after the burst) favor a scenario where
the optical light follows a pure power-law decay with an index of
$\alpha \sim -1.0$.  We find no evidence for a contribution from a
supernova like SN1998bw.  This suggests that either there are multiple
classes of long-duration gamma-ray bursts, or that the peak luminosity
of the supernova was $> 1.5$ mag fainter than SN1998bw.  The \hst\
images and EFOSC2 spectra indicate that the gamma-ray burst was
located in a bright, extended feature (possibly a star-forming region)
1.4 kpc from the nucleus of a $0.2 L^{\ast}_B$ galaxy at $z= 0.434$,
possibly a Seyfert~2 galaxy.  The late-time afterglow and host galaxy
of GRB~990712 bear some resemblance to those of GRB~970508.

\end{abstract}


\section{Introduction}

	The spatial association of GRB~980425 with the
unusual Type Ib/c supernova SN1998bw at $z=0.0085$ provided the first
tantalizing evidence that some gamma-ray bursts (GRBs) are related to
the end-stages of the lives of massive stars \citep{GVV1998}.  Recent
evidence for similar supernova (SN) signatures in the late ($\sim
15(1+z)$ days) light curves of the genuine cosmological GRBs
GRB~970228 \citep{D1999,R1999,GNV1999} and
GRB~980326 \citep{CTG1999,BKD1999} indicates that at
least some long-duration GRBs are related to SN explosions.  A GRB--SN
association suggests that the progenitors of GRBs are short-lived and
that GRBs die where they were born---in the star-forming regions of
their host galaxies \citep{P1998}.

	\citet{HH1999} found evidence from {\sl Hubble Space
Telescope\/} ({\hst}) Space Telescope Imaging Spectrograph (STIS)
imaging for a spatial coincidence between GRB~990123 and
a star-forming region in its host galaxy. The association of GRBs with
star-forming regions is important for models of their progenitors and
can be used to probe the physics of star formation and the global
star-formation history of the Universe \citep{MM1998,T1999,BN1999}.

	GRB~990712 was first localized by {\sl
BeppoSAX\/} and detected as having the strongest $X$-ray afterglow
observed to date \citep{HIT1999}.  \citet{BSM1999} discovered a
bright, decaying optical afterglow (OA) ($R = 19.4 \pm 0.1$) four
hours after the burst.  \citet{GVR1999} measured a preliminary
redshift of $z = 0.430\pm0.005$ from a set of absorption and emission
lines, which makes it the nearest GRB with a secure redshift that has
been observed to date (apart from SN1998bw).  ESO New
Technology Telescope images obtained 3.7 days after the burst led
\citet{HCC1999} to hypothesize the existence of a bright host galaxy
with $R = 22$ on the grounds of an apparent leveling off of the light
curve relative to the suspected power-law decline ($\alpha = -1.05$)
\citep{KH1999} of the OA\@.  Subsequent ESO Very Large Telescope
imaging \citep{HFD1999} confirmed the leveling off of the light curve,
and yielded evidence for the existence of an extended object
contributing to the flux at the position of the GRB\@.
\citet{HFD1999} predicted that the existence of a SN would lead to a
bump in the light curve around 1 August 1999, and that a SN model
could be distinguished from a no-SN model in late \hst\ and
ground-based imaging, as the OA would be brighter and the host fainter
in the SN scenario than in the no-SN scenario.  These predictions are
presented in \citet{S1999}, which reports the discovery and early
light curve of the OA of GRB~990712.

	In this {\paper} we present late {\hst} imaging, as well as
ground-based imaging and spectroscopy, aimed at testing these
predictions.  At the time of the {\hst} observations the $R$-band
magnitudes of the host galaxy and the OA are predicted to be
$22.25\pm0.05$ and $<23.91\pm0.05$ in the SN scenario, and
$21.75\pm0.05$ and $25.39\pm0.1 $ in the no-SN scenario.  We assume a
standard Friedman cosmology with $H_0 = 65\ \ho$, $\Omega_0 = 0.2$,
and $\Lambda = 0$.  At $z = 0.4337$ this corresponds to a scale of 5.6
proper kpc per arcsecond, a luminosity distance of 2.37 Gpc, a
distance modulus of 41.88, and a look-back time of 4.9 Gyr.  Including
a cosmological constant of $\Lambda = 0.8$ increases these values by
$\sim 10$\%.

\section{Late Ground-Based Imaging\label{SECTION:ground}}

	Direct images were obtained with the DFOSC on the Danish
1.54-m telescope at La Silla on 7 ($R$ band) and 9 ($V$ band) October
1999 UT, i.e., 87 and 89 days after the burst.  Exposure times were $3
\times 15$ minutes in each band and the seeing full-width at
half-maximum (FWHM) was $1\farcs8$.  The photometry was carried out
using {\sc SExtractor} v2.0.13 \citep{BA1996}.  The calibration was
tied to the internal reference stars of \citet{S1999} and gave $R =
21.92 \pm 0.08$ and $V = 22.40 \pm 0.08$ for the galaxy at the
location of the GRB\@.  Two more $R$ images (120 sec and 180 sec) were
obtained with the ESO 3.6-m telescope on 12 November 1999 UT (123 days
after the burst) in $0\farcs7$ seeing (see \S\ref{SECTION:spectra}).
These images yielded $R=21.91 \pm 0.05$ for the host galaxy.  The
ground-based images thus showed no signs of a transient source.  The
photometry is consistent with host magnitudes of $R=21.91\pm 0.04$ and
$V = 22.40\pm 0.08$.

\section{HST Imaging\label{SECTION:hst}}

\subsection{The STIS Data\label{SECTION:stis}}

        {\hst} observations of the OA were made on 29 August 1999 UT,
47.7 days after the burst, as part of the Cycle~8 program GO-8189.
Six 620 second exposures were taken with the STIS in each of its 50CCD
(clear, hereafter referred to as CL) and F28X50LP (long pass,
hereafter referred to as LP) modes.  The CCD gain was set to 1
e$^-$/ADU, giving a read-out noise of 4 e$^-$/pixel, and the data was
processed through the standard STIS pipeline.  We retrieved this data
from the {\sl HST\/} Data Archive and combined the images using the
{\sc Dither} (v1.2) software \citep{FH1999} as implemented in
IRAF\footnote{Image Reduction and Analysis Facility (IRAF), a software
system distributed by the National Optical Astronomy Observatories
(NOAO).}  (v2.11.1)/STSDAS (v2.0.2).  We used ``pixfrac'' $= 0.5$, and
a final output scale of $0\farcs0254$/pixel.
Figure~\ref{FIGURE:galaxy} shows the drizzled CL image of the probable
host galaxy.

	Conversion from counts to $AB$ magnitudes was achieved using
the zero points given in the STIS Instrument Handbook.  We assumed
that the galaxy had a power-law spectrum of the form $F_{\nu}(\nu) = k
\nu^\beta$, where $k$ is constant, and converted the $AB$ magnitudes
to standard Johnson $V$- and Kron-Cousins $R$-band magnitudes using
\begin{equation}
M = m_{\rm CL} + K_M + 48.6 
    + 2.5\beta\log_{10} \left(\nu_{\rm CL} / \nu_M\right),
\label{EQUATION:mag}
\end{equation}
where $M$ represents the $V$- or $R$-band magnitude, as appropriate,
$m_{\rm CL}$ and $m_{\rm LP}$ are the instrumental $AB$ magnitudes
measured in the CL and LP filters, $K_M$ is the appropriate zero
point, $\nu_{\rm CL}$ and $\nu_{\rm LP}$ are the central frequencies
of the CL and LP filters, and $\beta = - 0.4(m_{\rm CL}-m_{\rm LP}) /
\log_{10}\left(\nu_{\rm CL} / \nu_{\rm LP}\right)$.

\subsection{Photometry of the Optical Afterglow\label{SECTION:OA_mag}}

\begin{figure}
\resizebox{\hsize}{!}{\includegraphics{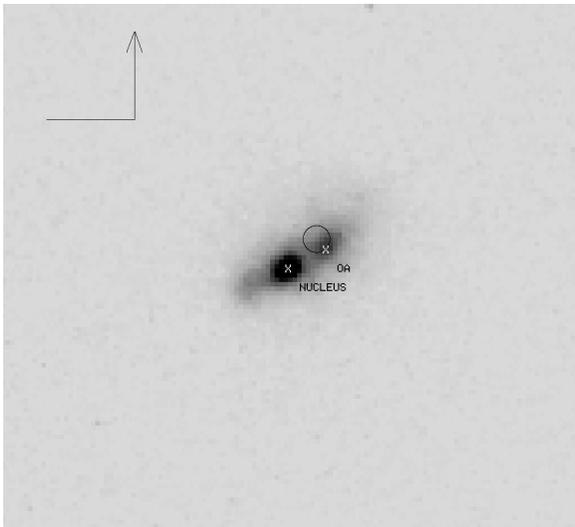}}
\caption{This figure shows the central $3\farcs25 \times 3\farcs25$
section of the drizzled 50CCD (clear) {\sl HST\/}/STIS image.  The
scale is $0\farcs0254$ per pixel.  The arrow indicates north and the
line indicates east.  Both lines are $0\farcs5$ long.  The circle
shows the error circle from matching the coordinates of the OA given
in \protect\citet{S1999} to the CL image.  The OA is the only point
source in this error circle.}
\label{FIGURE:galaxy}
\end{figure}

        We estimated the total $AB$ magnitude of the OA on both the CL
and the LP images by using the {\sc DaoPhot/AllStar II} \citep{S1987}
photometry package.  A point-spread function (PSF) was constructed in
the manner described in \S4 of \citet{HH1999}.  The OA was identified
by matching the coordinates of the OA given by \citet{S1999} to the CL
and LP STIS images, and looking for point sources inside the
\citet{S1999} error circle (see Fig.~\ref{FIGURE:galaxy}). The {\sc
AllStar} ``sharp'' statistic is related to the angular size of an
object relative to a PSF\@.  Isolated point sources will have
``sharp'' values of $\sim 0$ while extended sources will have
``sharp'' values $\gg 0.1$.  The only point source (``sharp'' $=
0.000$ in the CL filter and $0.053$ in the LP filter) on the STIS
images that lies within the error circle is located $0\farcs033$ ($=
1.3$ drizzled STIS pixels) from the \citet{S1999} position for the
OA\@.  Therefore, we conclude that this point source is the OA for
GRB~990712. The brighter, point-like feature 0\farcs24 to
the southeast of the OA has ``sharp'' $= 0.990$ in the CL filter which
suggests that it is the nucleus of the galaxy and not a point source
(see \S~\ref{SECTION:host}).

	Aperture corrections were performed by using the {\sc DaoPhot
II} {\sc addstar} routine to generate an artificial star with the same
instrumental magnitude as the OA and measuring the total flux in
apertures with radii of $1\farcs108$ (CL) and $0\farcs963$ (LP).
Tables~14.3~and~14.5 of the STIS Instrument Handbook suggest that
these radii correspond to $100$\% of the encircled energy in the
PSF\@.  The OA magnitudes are $V = 25.69 \pm 0.02$ and $R = 25.23 \pm
0.09$, which yields a color of $V\!-\!R = 0.46 \pm 0.09$.  The
Galactic reddening towards the OA ($b^{\rm II} = -40\fdg19$, $l^{\rm
II} = 315\fdg28$) is $E_{B\!-\!V} = 0.033$ \citep{SFD1998}.  We used
$A_V = 0.11$ and $A_R = 0.08$ to obtain extinction-corrected
magnitudes of $V_0 = 25.58 \pm 0.02$ and $R_0 = 25.15 \pm 0.09$.

	A visual examination of each STIS image, after the PSF for the
OA has been subtracted, shows that we have slightly oversubtracted the
OA\@.  We estimate that this oversubtraction has resulting in our
magnitudes being overestimated by at most $0.23$ mag in the $V$ band
and $0.40$ mag in the $R$ band.  Therefore, the true magnitudes of the
OA are $V_0 = 25.58$--$25.81$ and $R_0 = 25.15$--$25.55$.  The
corresponding range in color is ${(V\!-\!R)}_0 = 0.26$--$0.43$, and in
spectral index is $-1.24 \le \beta_{\rm OA} \le -0.37$.

\begin{figure}
\resizebox{\hsize}{!}{\includegraphics{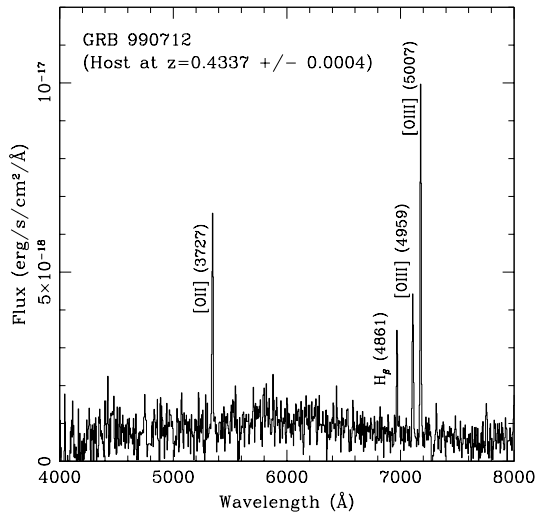}}
\caption{This figure shows the EFOSC2 spectrum of the host galaxy,
obtained 123 days after the burst. The absolute flux calibration has
been derived from our $V$-band photometry of the galaxy. The spectral
resolution is $R\sim550$ ($4.28$ {\AA} per pixel). The line widths and
ratios indicate that the host may be a Seyfert~2 galaxy.}
\label{FIGURE:spectrum}
\end{figure}

\section{Spectroscopy of the Host\label{SECTION:spectra}}

	We used the ESO Faint Object Spectrograph and Camera 2
(EFOSC2), mounted at the cassegrain focus of the ESO 3.6-m telescope,
to obtain a 30-minute optical spectrum of the host galaxy of
GRB~990712 on 12 November 1999 UT\@.  The weather
conditions were good, with $0\farcs7$ seeing and photometric sky.  We
observed in binned mode, with an effective pixel size of $0\farcs32$
and a resolution of $4.28$ {\AA}/pixel (or $R \sim 550$) between 4000
{\AA} and 8000 {\AA}.  The slit was $1\farcs2$ wide and oriented along
the parallactic angle.  The data were reduced with IRAF and the CTIO
long-slit package.  A spectrum of the standard star
LTT9491 \citep{L1992} was used to perform the relative
flux calibration.  As the standard was observed with a relatively
narrow slit ($1\farcs2$), we tied the absolute flux calibration to our
measurement of the $V$-band flux for the galaxy.

	The spectrum (Fig.~\ref{FIGURE:spectrum}) exhibits four
prominent narrow (unresolved) emission lines which we identify as
[\ion{O}{2}]$\lambda$3727, H${\beta}\,\lambda$4861, and
[\ion{O}{3}]$\lambda\lambda$4959,5007.  Table~\ref{TABLE:lines} lists
the observed- and rest-frame wavelengths, $\lambda$ and $\lambda_0$,
the redshifts, $z$, the observed- and rest-frame line widths, $W$ and
$W_0$, and the line fluxes, $f$, for each line.  The mean redshift
derived from these lines is $z = 0.4337 \pm 0.0004$ where the
uncertainty is the $3\sigma$ uncertainty in the mean.  We detect no
trace of stellar absorption features such as the Ca H and K
($\lambda\lambda$3968,3933) lines or the 4000 {\AA} break, and we do
not detect any obvious intervening absorption systems along the line
of sight.

\begin{deluxetable}{clllllc}
\tablewidth{0pt} 
\tablecaption{Emission line identifications and fluxes\label{TABLE:lines}}

\tablehead{%
\colhead{Line} &
\colhead{$\lambda$}&
\colhead{$\lambda_0$} &
\colhead{$z$} &
\colhead{$W$}&
\colhead{$W_0$}&
\colhead{$f$} \\
\colhead{} &
\colhead{(\AA)} &
\colhead{(\AA)} &
\colhead{} &
\colhead{(\AA)} &
\colhead{(\AA)} &
\colhead{(10$^{-17}$ erg s$^{-1}$ cm$^{-2}$)} 
}
\startdata
[\ion{O}{2}]   &  5345.48 &  3727.4 & 0.4341 & $66\pm10$  & $46\pm7$   &  6.5 \\
 H$\beta$      &  6969.20 &  4861.3 & 0.4336 & $30\pm5$   & $21\pm3.5$ &  2.5 \\
{[\ion{O}{3}]} &  7108.68 &  4958.9 & 0.4335 & $75\pm10$  & $52\pm7$   &  4.6 \\
{[\ion{O}{3}]} &  7177.80 &  5006.9 & 0.4336 & $195\pm30$ & $136\pm21$ & 10.9 \\
\enddata
\end{deluxetable}

	We note that the [\ion{O}{3}]$\lambda$5007 line lies near at
the red edge of the bandpass of the $R$ filter.  This may complicate
the interpretation of late-time $R$-band light curves of the OA when
the data is collected using different instruments, and when the total
flux is dominated by the light of the host galaxy.

\section{The Host Galaxy\label{SECTION:host}}

	The host galaxy is an extended source with elliptical
isophotes, a bright, concentrated nucleus, and a bright, extended
feature to the northwest of the nucleus.  The nucleus has a FWHM of
$0\farcs11$ ($= 0.62$ kpc) in the CL filter, which is only slightly
wider than the PSF\@.  There is some indication that the isophotes
twist to the north at the northwest end of the galaxy and to the south
at the southeast end.

	The integrated $V$- and $R$-band magnitudes of the galaxy were
obtained by subtracting the light from the OA and performing aperture
photometry with an aperture of radius $2\farcs5$ in each of the CL and
LP images.  The measured spectral index is $\beta_{\rm gal} = -2.69$,
and Eq.~\ref{EQUATION:mag} yields $V = 22.51 \pm 0.04$ and $R = 21.80
\pm 0.06$.  Correcting for Galactic extinction, and assuming no
internal extinction in the host galaxy, gives ${(V\!-\!R)}_0 = 0.68
\pm 0.07$.  For $z = 0.4337$ the observed $R$-band is approximately
equivalent to the rest-frame $B$-band, so $M_B \sim -19.5$.
\citet{LTH1995} finds $M^{\ast}_B = -21.23$ in the rest frame for blue
galaxies with redshifts of $0.2 \le z \le 0.5$.  This corresponds to
the host galaxy for GRB~990712 having $L_B \sim 0.2
L^{\ast}_B$ where $L^{\ast}_B$ is the $B$-band luminosity of a typical
blue galaxy at $z = 0.4337$.

	We estimated the star-formation rate (SFR) in the host galaxy
using Eq.~2 of \citet{MP1998} as described in \citet{HH1999}.  The
total SFR is $0.29$--$0.45 {\cal M}_{\sun}$ yr$^{-1}$, depending on
the assumed initial mass function (IMF).  The SFR is corrected for
extinction in our Galaxy, but it assumes that there is no dust, or
obscured star formation, in the host galaxy.  This is probably a poor
assumption so our derived SFR should be considered to be a lower limit
on the true SFR in the host.  The implied [\ion{O}{2}] luminosity
(Table~\ref{TABLE:lines}), corrected for Galactic extinction, is
$L_{3727} = 6.3\times 10^{40}$~erg~s$^{-1}$.  If we assume that the
strength of the [\ion{O}{2}] line is related to star formation then
this corresponds to a SFR of $0.88\pm0.25 {\cal M}_{\sun}$ yr$^{-1}$
\citep{K1998}. This is 2--3 times as large as the SFR derived from the
continuum flux, possibly indicating internal extinction in the host or
a contribution from non-thermal emission. The derived SFR (from the
[\ion{O}{2}] flux) is $\sim 20$\% of the SFR found by \citet{BOD1999}
for the host of GRB~990123 and comparable to that of the
host of GRB~970508 \citep{BDK1998}.  The specific SFR per
unit luminosity of the GRB~990712 host galaxy is $\sim
0.4$ times that of the host galaxies of GRB~990123 and
GRB~970508.

	The OA is located in the bright, extended source $0\farcs242$
($= 1.4$ kpc) to the northwest ($-50\arcdeg$ north of east) of the
nucleus (see Fig.~\ref{FIGURE:galaxy}).  This region has a FWHM of
$0\farcs28$ ($= 1.6$ kpc), which is significantly more extended than
the STIS PSF (FWHM $= 0\farcs09$).  The surface brightness, after
subtracting the OA and correcting for Galactic extinction, is
$\mu_{V,0} = 15.83 \pm 0.01$ and $\mu_{R,0} = 15.04 \pm 0.01$ for an
integrated color of ${(V\!-\!R)}_0 = 0.79 \pm 0.01$.  The OA is
located $\sim 0\farcs025$ ($\sim 140$ pc) south of the center of this
extended structure.  The total $V$-band flux in the feature, after
subtracting the flux from the [\ion{O}{2}] emission line, is $0.323
\pm 0.003$ $\mu$Jy.  Assuming a power-law spectrum with $\beta =
-2.93$ the SFR is $0.03$--$0.05 {\cal M}_{\sun}$ yr$^{-1}$ depending
on the IMF\@.  Again, we wish to stress that this is a lower limit on
the true SFR in the feature.  The estimated SFR is approximately one
third of the SFR that \citet{HH1999} found for the star-forming region
that coincides with the position of GRB~990123 whereas
the diameters and luminosities are comparable.

	The spectroscopic data demonstrate that all emission lines,
both forbidden and permitted, have narrow widths that are consistent
with a Seyfert~2 galaxy (the instrumental resolution only gives an
upper limit on the rest-frame velocity widths of a few hundred
$\kms$).  The ratio [\ion{O}{3}]/H$\beta$ is greater than three, which
is indicative of an active galaxy \citep{SO1981}.  The object lies on
the borderline between \ion{H}{2} regions and narrow-line AGNs in the
$\log($[\ion{O}{3}]$/{\rm H}\beta)$ vs.\
$\log($[\ion{O}{2}]/[\ion{O}{3}]) diagram---uncorrected for
extinction, see \citet{BPT1981}.  This suggests that the host galaxy
of GRB~990712 may be a Seyfert~2 galaxy, but further
spectroscopic observations will be needed to confirm this.  For
example, a measurement of $\log($[\ion{O}{1}]$/{\rm H}\alpha)$ from
near-IR spectroscopy would discriminate between a star-forming galaxy
or a Seyfert~2 \citep{VO1987}.

\section{The Late Afterglow and the Possible GRB~990712--Supernova 
Association}

\begin{deluxetable}{cccc}
\small
\tablewidth{0pt}
\tablecaption{Predicted and observed magnitudes\label{TABLE:predictions}}

\tablehead{%
\colhead{Quantity}   &
\colhead{PL}       &
\colhead{SN+PL}    &
\colhead{Observed}
}
\startdata
$V_\mathrm{OA}$             & $25.64\pm0.1$  & $<25.16\pm0.05$ & $25.81\pm0.12$ \\
$R_\mathrm{OA}$             & $25.39\pm0.1$  & $<23.91\pm0.05$ & $25.43\pm0.20$ \\
${(V\!-\!R)}_\mathrm{OA}$   &  $0.25\pm0.1$  & $\sim1.25$      &  $0.38\pm0.09$ \\
$V_\mathrm{host}$           & $22.30\pm0.05$ & $22.70\pm0.05$  & $22.47\pm0.04$ \\
$R_\mathrm{host}$           & $21.75\pm0.05$ & $22.25\pm0.05$  & $21.88\pm0.04$ \\
${(V\!-\!R)}_\mathrm{host}$ &  $0.55\pm0.07$ & $0.45\pm0.07$   &  $0.63\pm0.10$ \\
\enddata

\end{deluxetable}

	Table~\ref{TABLE:predictions} summarizes the predicted and
observed $V$- and $R$-band magnitudes for the OA and its host galaxy
in both the pure power-law (PL), and power law + supernova (PL+SN)
scenarios.  The magnitudes for the host galaxy are the weighted means
of the magnitudes that we obtained from the Danish 1.54-m and ESO
3.6-m telescopes, and the {\sl HST}.  Allowing for small systematic
uncertainties, the observed magnitudes and colors for the host and OA
are consistent with the predictions of the pure power-law model, which
indicates that the OA followed a slow power-law decay with a constant
index of $\alpha \sim -1.0$, with no significant late-time break
towards steeper decay.  Our results are inconsistent with the presence
of a SN like SN1998bw.  The discrepancy between the
observations and the predictions of the PL+SN scenario suggests that a
SN in the late-time light curve of GRB~990712 would have
had to be $> 1.5$ mag fainter than SN1998bw to be
consistent with the data presented in this {\paper}.  Therefore, we
conclude that either GRB~990712 did not produce a SN, or
that the flux received from the SN was much smaller than expected from
scaling SN1998bw to $z = 0.4337$.

	If there was no SN then this supports the view that
long-duration GRBs have more than one type of progenitor
\citep{LW1999}.  GRB~970508 is the only other GRB for
which there is some evidence that a SN was {\sl not\/} present.
However, the paucity of observations around the time of the predicted
peak luminosity of the SN makes it difficult to unambiguously rule out
a SN contribution in GRB~970508's light curve.
Therefore, GRB~990712 provides the first solid evidence
that not all long-duration GRBs are associated with standard-candle
SNe.  There are some similarities between GRB~990712 and
GRB~970508 that suggest that they may be members of the
same class of bursts.  Both had shallow late-time decay slopes
($\alpha \sim -1.0$), both appear to lack a 1998bw-type SN, both
bursts were strong $X$-ray sources, and both bursts have host galaxies
that are significantly fainter than $L^{\ast}$.

	If there was a SN associated with GRB~990712,
then its peak intensity was $> 1.5$ mag fainter than that of
SN1998bw.  Type Ib/c SNe are poor standard candles since
their predicted peak intensities can vary by 1--2 mag, with a mean
peak $B$-band magnitude that is 1.16 mag fainter than that of
SN1998bw \citep{VBT1991}.  Moreover, the predicted peak
magnitude for a SN associated with GRB~990712 depends on
the cosmological model, the exact spectral shape of
SN1998bw at very short wavelengths, the width of the
lightcurve peak, the possible extinction in the host galaxy, and the
possible evolution of SNe with redshift. A large sample of GRB
afterglows with measured redshifts and detectable SN signatures is
needed to establish the intrinsic luminosity distribution of SNe
accompanying GRBs.

\acknowledgments

FC is supported by Chilean grant FONDECYT/3990024. Additional funding
from the European Southern Observatory is gratefully
acknowledged. This work was supported by the Danish Natural Science
Research Council (SNF).

\end{document}